\documentclass{article}

\PassOptionsToPackage{round}{natbib}

\usepackage{neurips_2020}
\usepackage{pslatex}
\usepackage{hyperref}
\usepackage{lineno}
\usepackage{xcolor}
\usepackage{bookmark}
\usepackage{graphicx}
\usepackage{tikz}
\usepackage{amsmath}
\usepackage{amssymb}
\usepackage{amsthm}
\usepackage{mathtools}
\usepackage{placeins}
\usepackage{float}
\usepackage{bbm}
\usepackage[capitalize,noabbrev]{cleveref}
\usepackage{enumitem}
\usepackage{svg}
\usepackage{booktabs}
\usepackage{changes}
\usepackage{wrapfig}
\usepackage{caption}

\usepackage{amsmath,amsfonts,bm}

\def\eqref#1{equation~\ref{#1}}

\def\1{\bm{1}}

\def\mX{{\bm{X}}}
\def\mY{{\bm{Y}}}
\def\mZ{{\bm{Z}}}

\DeclareMathAlphabet{\mathsfit}{\encodingdefault}{\sfdefault}{m}{sl}
\SetMathAlphabet{\mathsfit}{bold}{\encodingdefault}{\sfdefault}{bx}{n}

\title{
    Optimizing fMRI Data Acquisition for Decoding Natural Speech with Limited Participants
}

\author{
    Louis~Jalouzot\\
    CEA, ENS\\
    Université Paris-Saclay\\
    France\\
    \scriptsize{\texttt{jalouzot.louis@gmail.com}}\\
    \And
    Alexis~Thual\\
    karavela.ai\\
    France
    \And
    Yair~Lakretz\\
    ENS, EHESS, CNRS\\
    Université PSL\\
    France
    \And
    Christophe~Pallier\\
    INSERM, CEA, CNRS\\
    Université Paris-Saclay\\
    France
    \And
    Bertrand~Thirion\\
    INRIA, CEA\\Université Paris-Saclay\\
    France
}

\begin{document}

\maketitle

\begin{abstract}
    We investigate optimal strategies for decoding perceived natural speech from fMRI data acquired from a limited number of participants. Leveraging \cite{lebel_natural_2023}'s dataset of 8 participants, we first demonstrate the effectiveness of training deep neural networks to predict LLM-derived text representations from fMRI activity.
    Then, in this data regime, we observe that multi-subject training does not improve decoding accuracy compared to single-subject approach.
    Furthermore, training on similar or different stimuli across subjects has a negligible effect on decoding accuracy.
    Finally, we find that our decoders better model syntactic than semantic features, and that stories containing sentences with complex syntax or rich semantic content are more challenging to decode.
    While our results demonstrate the benefits of having extensive data per participant (deep phenotyping), they suggest that leveraging multi-subject for natural speech decoding likely requires deeper phenotyping or a substantially larger cohort.
\end{abstract}

\textbf{Keywords:} fMRI; decoding; natural speech; deep learning

\section{Introduction}

Early on, the field of neuroscience has been interested in decoding percepts from recorded brain activity.
Using functional MRI (fMRI), \cite{Kamitani_Tong_2005} paved the way for decoding visual stimuli through retinotopy,
while \cite{formisanoWhoSayingWhat2008} could identify individual words or short phrases.
More recently, several publications have tackled similar challenges with increasing decoding accuracy.
These improvements can in part be attributed to the acquisition of deep-phenotyping datasets, where a large amount of data is acquired in each participant.
In particular, \cite{ozcelik_natural_2023} trained accurate decoders of visual semantics using the Natural Scenes Dataset \citep{allen_massive_2022},
and \cite{tang_semantic_2023,ye2025generative} could decode natural language from the LeBel dataset \citep{lebel_natural_2023}.\\
However, one major challenge to leverage data acquired from multiple participants is inter-subject variability.
Although fine-tuning \citep{scotti_mindeye2_2024} or functional alignment \citep{thual_aligning_2023} can partially address this issue, it remains unclear how to train decoders that model this variability.
The most recent approaches are based on linear mappings between feature spaces and subject data \cite{daiMindAlignerExplicitBrain2025}.

In this work, we focus on the specific case of decoding representations of perceived natural speech from deep-phenotyped fMRI data.
We implement a decoding-first approach, where we directly predict text features from recorded brain activity using methods from visual perception decoding.
Indeed, recent publications hint that, regardless of the input (M/EEG, fMRI, etc) and output (vision, language, etc) modalities,
current state-of-the-art decoders are typically obtained by training a deep neural network on a contrastive objective \citep{defossez_decoding_2022,scotti_reconstructing_2023,dascoliDecodingIndividualWords2024}.\\
Besides, we seek to derive insights on how to optimize data acquisition strategies for decoding natural speech when a limited number of participants are available.

\newpage

We make the following contributions:
\begin{enumerate}
    \item We demonstrate the effectiveness of training deep neural networks to predict LLM-derived text representations from fMRI activity using a contrastive objective.
    \item We find that decoding performance scales with the quantity of training data available per participant, and that it does not plateau even with 13.5 hours of training data.
    \item We show that training decoders on multiple subjects does not improve decoding accuracy compared to single-subject approaches in our current data regime (high amount of data per participant, but low number of participants).
    \item Moreover, we find that stimulus overlap between subjects has a negligible effect on decoding accuracy in multi-subject setups.
    \item Finally, we show that current language decoders better model syntactic than semantic features, and that sentences with complex syntax or rich semantic content are more challenging to decode.
\end{enumerate}

In terms of experimental design, our work indicates that, for decoding, it is more effective to acquire a large amount of data per participant (deep phenotyping) than acquiring data from multiple participants.
This conclusion holds at least in the current data regime, where we have a large amount of data per participant (13.5 hours), but only eight participants.

Our approach differs from that of \cite{tang_semantic_2023}: the authors generate text using a beam search algorithm which selects the most likely word at each step, evaluating its likelihood using a pre-trained encoding model predicting brain activity from text.
Another approach closer to our work is presented in \cite{ye2025generative} where they train a decoder to predict embeddings from fMRI to feed a prompt to an LLM to regenerate the perceived text.

\section{Methods}

\begin{figure*}
    \centering
    \includegraphics[width=\textwidth]{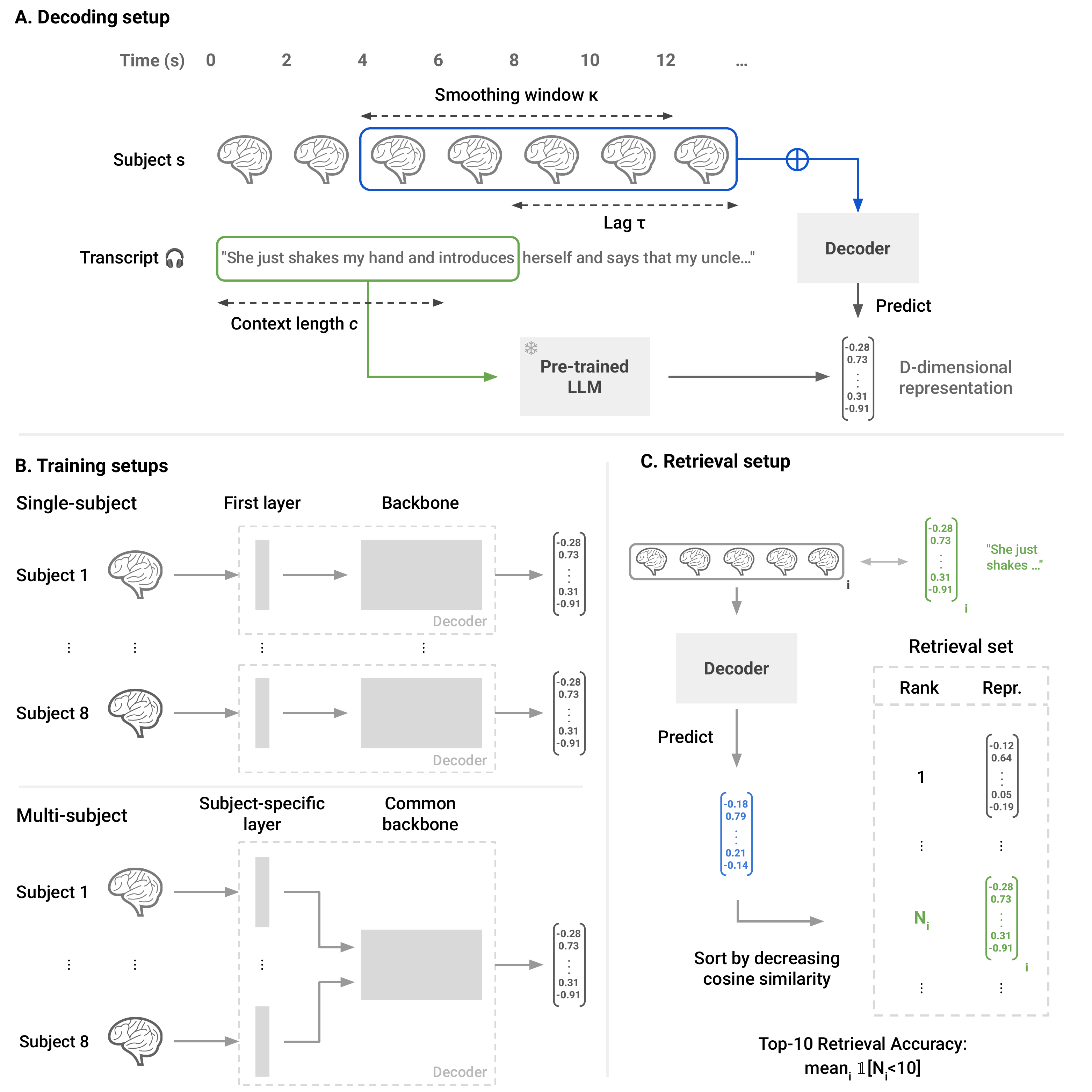}
    \vspace{0.3cm}
    \captionsetup{margin=5pt}
    \caption{
        \textbf{Method for decoding natural speech from fMRI activity}\\
        \textbf{A. Decoding setup} Deep Neural Networks are trained with a contrastive objective to predict text representations (derived from Large Language Models embeddings) from fMRI activity recorded as participants listened to natural speech.
        Key parameters include \textit{context length} $c$, the number of prior chunks added to the text representations, \textit{lag} $\tau$, the delay between neural activity and the hemodynamic response, and \textit{smooth} $\kappa$, the number of preceding brain volumes averaged.\\
        \textbf{B. Subject approaches} We compare single-subject (one decoder per subject) and multi-subject (shared decoder backbone with subject-specific layers at the bottom) approaches.\\
        \textbf{C. Retrieval setup} Decoders are evaluated in a retrieval setup, we rank chunks from a retrieval set (candidates) by the cosine similarity between their representation and the predicted one. Then we compute top-10 accuracy (frequency of the ground truth appearing among the top 10 candidates).
    }
    \label{fig:method}
\end{figure*}

We seek to train models predicting text representations (LLM embeddings) from fMRI activity, using a contrastive objective.

\subsection{General setup}

\paragraph{fMRI data}
We rely on the \cite{lebel_natural_2023}'s dataset which comprises recordings from eight participants as they listened to a rich array of 27 stories from the \emph{Moth Radio Hour}, encompassing $\sim$~6 hours of auditory stimuli per individual.
Furthermore, 3 of those subjects listened to an additional 57 stories, resulting in 16.5 hours of data for these participants.
The fMRI data acquisition was performed at 3T, employing a spatial resolution of 2.6mm isotropic voxels and a temporal resolution of 2 seconds (TR=2s).

\paragraph{Decoding task}
Our primary objective is to train models capable of decoding text representations directly from fMRI activity.
For each participant $s$, we denote the fMRI activity as $(\mX^s_{t,n})_{t,n} \in \mathbb{R}^{T, N}$, where $T$ represents the number of brain volumes acquired and $N$ is the total count of voxels.
Simultaneously, we have the transcripts $(\mZ^s_{t})_{1 \leq t \leq T}$ of the text heard by the participant during the acquisition of each brain volume.
Using a pre-trained large language model on the transcripts, we derive $D$-dimensional text representations, or \textit{chunks}, denoted as $\mY^s_{t} \triangleq h(\mZ^s_{t}) \in \mathbb{R}^{T, D}$, where $h$ represents the embedding function of the Large Language Model (LLM).
The decoding task consists in learning a mapping from fMRI signals to these high-level text representations.
A direct approach to decoding would be to find a function $f$ that accurately predicts the text embedding $\mY^s_{t}$ from the simultaneously acquired fMRI volume $\mX^s_{t}$, such that $\hat{\mY}^s_{t} \triangleq f(\mX^s_{t}) \approx \mY^s_{t}$.
However, to account for the inherent delay between neural activity and the hemodynamic response \citep{ogawa_brain_1990}, we introduce a \textit{lag} parameter $\tau$.
This parameter shifts the prediction target in time, such that we aim to predict the text embedding $\mY^s_{t}$ from the brain volume acquired at a future time point $t + \tau$.\\
Since the transcript corresponding to a single brain volume may be short or even empty, text representations are computed from the concatenation of matching and preceding transcripts.
The number of preceding transcripts is denoted as $c$ and referred to as the \textit{context length}, such that $\mY^s_{t} \triangleq h(\mZ^s_{t-c:t})$.
This contextual window enriches the text representation with preceding linguistic context, potentially improving the decoder's ability to capture meaningful semantic information.

\paragraph{Preprocessing}
First we preprocess the fMRI data using \texttt{fmriprep} \citep{esteban_fmriprep_2019},
such that volumetric data from each participant is anatomically aligned to the MNI152 template \citep{mazziotta_probabilistic_1995}.
We subsequently apply a temporal smoothing technique.
Specifically, for each time point $t$, we averaged the current brain volume with the $\kappa$ preceding volumes, where $\kappa$ is the \textit{smoothing window} parameter.
This temporal averaging acts as a smoothing operation, aiming to reduce noise and stabilize the fMRI signal, thus potentially improving the signal-to-noise ratio for decoding purposes.
Following temporal smoothing, and using the \texttt{scikit-learn} library \citep{pedregosa_scikit-learn_2011}, we apply standard scaling to the activity of each voxel across the time dimension.
For the text embeddings, we use LLM2Vec \citep{behnamghaderLLM2VecLargeLanguage2024}.
These embeddings, with a hidden dimension of $D=4096$, were selected for their strong sentence representation capabilities, as they have been fine-tuned for this purpose from a Llama model.
Prior to decoder training, the text embeddings were normalized along the hidden dimension and then standard scaled across the time dimension, mirroring the preprocessing applied to the fMRI data.
Furthermore, to reduce the dimensionality of the input fMRI data and focus on the most informative voxels, we implement a voxel selection procedure based on encoding performance.  Specifically, we use Ridge regression with a regularization hyperparameter $\alpha=1$ to predict fMRI activity from the text embeddings on the training data, compute the $R^2$ score on a validation dataset for each voxel, and keep the top 4096 voxels with the highest scores. This step is crucial for managing computational complexity and ensuring that the decoder primarily learns from brain regions most relevant to speech processing.

\paragraph{Training}
We model the function $f$ from our decoding formulation using a deep neural network (DNN)
implemented in PyTorch \citep{neurips2019_9015}.
The architecture of this DNN incorporates 3 MLP layers, dropout, layer normalization and skip connections, inspired by successful architectures in related domains \citep[see e.g.][]{scotti_mindeye2_2024}.\\
We optimize this DNN for a contrastive loss \citep{radford_learning_2021} within a retrieval-based setup.
This approach is commonly employed in decoding to mitigate overfitting and enhance generalization, particularly when dealing with high-dimensional neuroimaging data and text representations in our case.\\
Model training incorporates early stopping, monitored on a validation set comprising 5\% of the data, to prevent overfitting.
Furthermore, the learning rate is reduced when the monitored objective plateaus.
Hyperparameters for the DNN architecture and training process are optimized through a Bayesian grid-search, leveraging \texttt{hydra} \citep{Yadan2019Hydra}, \texttt{submitit}\footnote{\url{https://github.com/facebookincubator/submitit}}, and \texttt{optuna} \citep{optuna_2019}, to maximize the cross-validated top-10 accuracy on a subset of subjects (subjects 1, 2, and 3) in the multi-subject setup.
The resulting optimized hyperparameters are displayed in \cref{table:hyperparameters}. Our code will be made available upon publication.

\paragraph{Evaluation}
We split the data into training, validation and test sets, and ensure that data from each of the three sets were acquired on different fMRI runs.
We evaluate the performance of our decoders on a retrieval task.
Given a validation (resp. test) brain volume $\mX^s_{t}$, the decoder predicts a text representation $\hat{\mY}^s_{t}$.
We compute the cosine similarity of $\hat{\mY}^s_{t}$ with each text embeddings in the validation (resp. test) retrieval set $\{ \mY^s_{k} \}_{k}$ and rank them.
Top-ranked chunks are called \textit{candidates}.
Our primary evaluation metric is top-10 accuracy, which is the proportion of instances where the ground truth text embedding is present among the top 10 candidates.
Note that the retrieval set consists of all text embeddings from the corresponding data split, and thus contains several thousand samples.\\
We employ cross-validation to obtain reliable performance estimates.
For subjects with less data, we use 5 folds, while for subjects with larger data, we use 15, ensuring that the test retrieval set consistently contains approximately 2000 chunks for each fold, thus maintaining comparability across subjects.
Across all performance plots, we visually represent the variability in decoding accuracy using shaded areas, which denote the 95\% confidence interval across the folds of cross-validation.
Moreover, we track two NLP similarity metrics.
First, we assess the semantic similarity between ground truth and retrieved text chunks using a GloVe bag-of-words cosine similarity restricted to content words (nouns, verbs, and adjectives), disregarding function words and structure.
Second, syntactic similarity is assessed using the Levenshtein edit distance \citep{Levenshtein_1965} between POS-tagged chunks, capturing structural resemblance of sentences in terms of grammatical categories.

\subsection{Additional setups}
\paragraph{Single-subject vs. Multi-subjects decoding}  In the \emph{Single-subject} setup, we train 8 independent decoders, one for each participant.
In the \emph{Multi-subject} setup, we train a single decoder on data from all subjects, but the first layer of the decoder is subject-specific, while the remaining parameters and the backbone of the decoder are shared across participants (see \cref{fig:method}.B).
We do not to \emph{explicitly} model inter-subject variability. Rather, we select voxels independently for each subject and the Subject-Specific Layers are used to project each subject's activation into a shared space, input of the backbone.

\paragraph{Effect of Stimuli overlap}
Is it important to train the multi-subject model on the same texts across subjects or not? To examine this question, we vary the proportion of stories shared during training -- from 0\% to 100\% -- while keeping the training set size constant.

\paragraph{Comparison of best vs.~worst decoded texts}

The podcasts or stories from the \emph{Moth Radio Hour} heard by the participants in the scanner varied quite a bit in content. We found that the decoding performance was much better for some than for others.
We perform a qualitative analysis using Gemini \citep{reidGemini15Unlocking2024} to analyse the commonalities  and differences between the stories with the best and the worst decoding performance (based the fMRI data from  subject 3.)

\section{Results}

\begin{figure*}[!t]
    \centering
    \begin{minipage}{0.49\textwidth}
        \includegraphics[width=\linewidth]{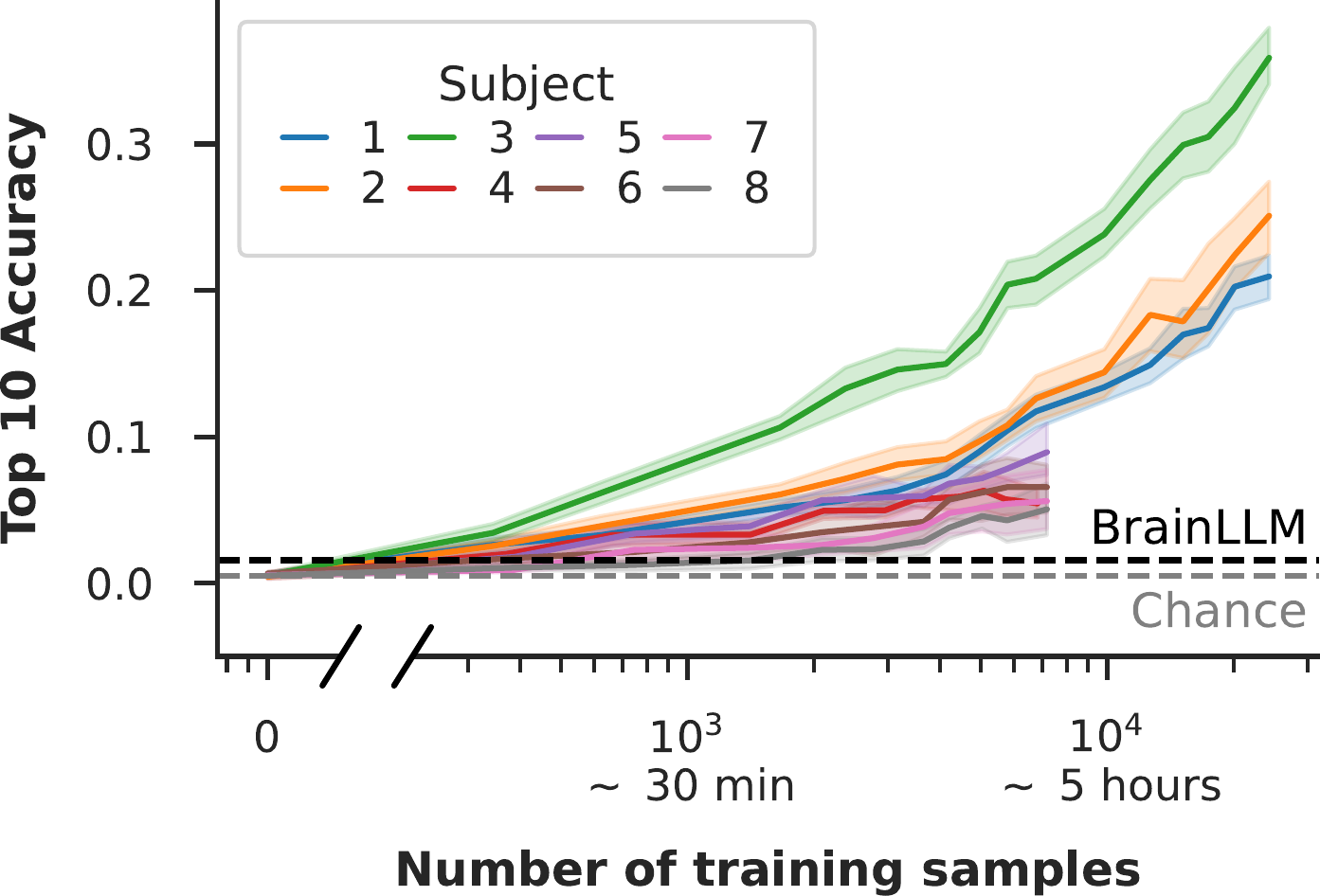}
        \vspace{0.1cm}
        \captionsetup{margin=5pt}
        \caption{
            \textbf{Impact of the amount of training data on single-subject performance} Cross-validated top-10 accuracy of single-subject decoders trained on varying amounts of data.
            The retrieval set contains about 2k samples, which were acquired on different MRI sessions than the training data, and come from different stories than that of the training set.
            We display chance level performance (0.05\%, grey) and the BrainLLM baseline (1.6\%, black) for comparison.
            Note the x-axis break to display chance-level performance of untrained decoders.
        }
        \label{fig:increase_data}
    \end{minipage}\hfill
    \begin{minipage}{0.49\textwidth}
        \includegraphics[width=\linewidth]{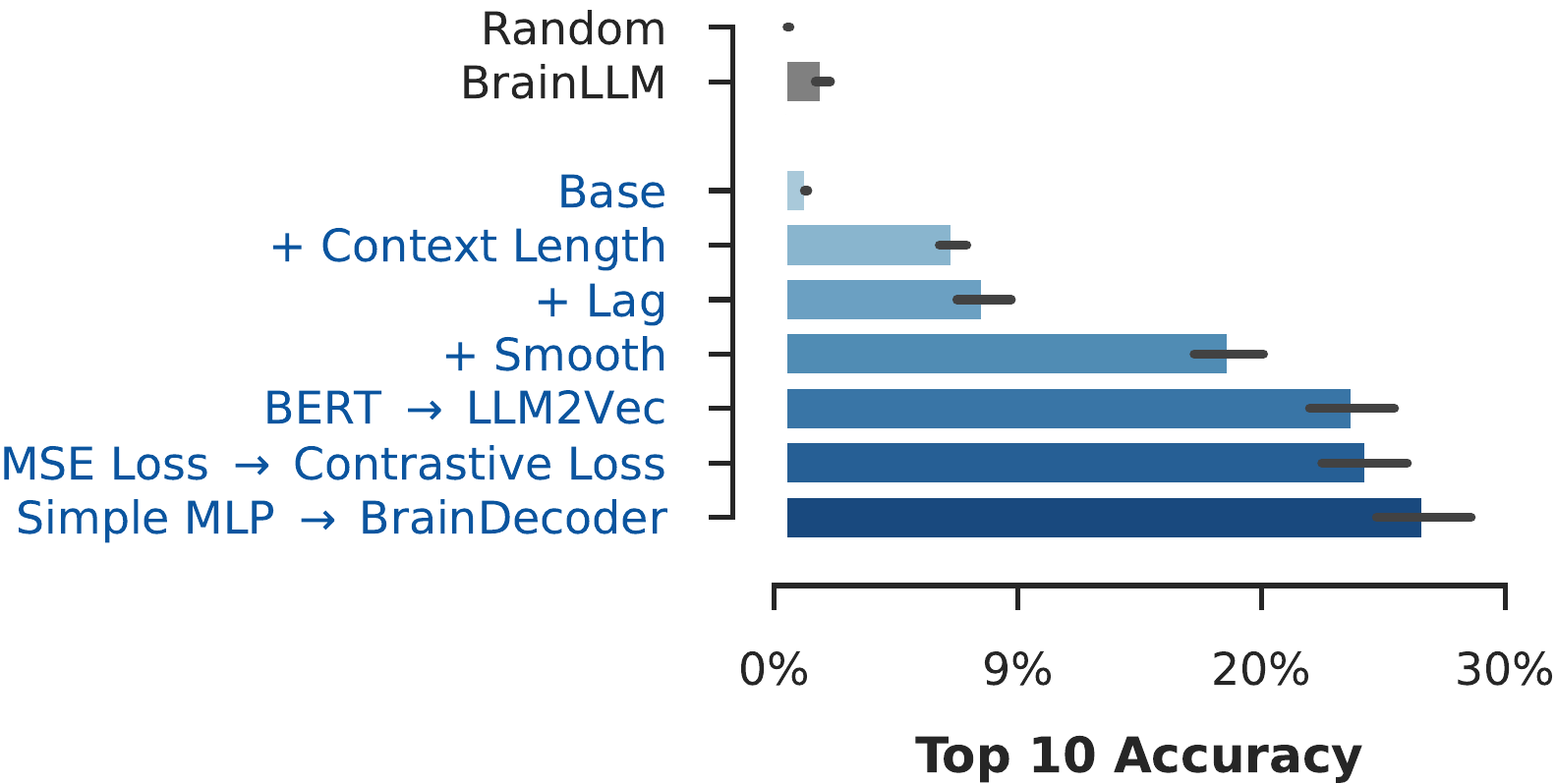}
        \vspace{0.1cm}
        \captionsetup{margin=5pt}
        \caption{
            \textbf{Setup comparison} Impact of various elements of the decoding setup on decoding performance. We start from a very crude version of our setup, namely "Base", which is essentially a simple MLP trained with MSE loss on BERT latents. Then each row corresponds to the previous setup with a modification described by its blue label. We display the top-10 accuracy obtained when training on subjects 1, 2 and 3 with SSLs and the full data. "Random" corresponds to a decoder that produces random representations, resulting in randomly ordered candidates in the retrieval set and thus achieving chance-level performance. We also display the BrainLLM baseline performance.
        }
        \label{fig:setup}
    \end{minipage}
\end{figure*}

\paragraph{Baseline}

The work closest to ours is that of \cite{ye2025generative}, who trained a decoder to predict fMRI embeddings to feed into a prompt to an LLM to regenerate the perceived text.
Although they have a different setup and training objective, they use the same dataset \citep{lebel_natural_2023} with the same chunking approach and a similar neural network architecture.
Therefore, their model constitutes an interesting baseline to compare our results to.
Originally, for a given chunk, they predict 4 token embeddings from 4 brain images and complete a prompt with the text preceding the chunk.
Then an LLM is used to generate the text.
To adapt these data to our evaluation setup, we average the 4 token embeddings and compute a \textit{ ground truth} as the embedding of the target text with the preceding text included in the prompt, averaged on the token dimension.
The authors kindly provided us with the predicted embeddings for the first 4 subjects on one test split.
The latter contains $\sim$~1.7k chunks so that the top 10 accuracies obtained can be compared to ours.
Even though their model has not been trained for retrieval, the predicted embeddings are still very informative as we obtained top-10 accuracies far above the chance-level of 0.05\% with 1.15\%, 1.66\%, 2.01\%, and 1.43\% for subjects 1, 2, 3, and 4 respectively. The average (1.6\%) is displayed as \textit{BrainLLM} baseline on Figures~2 and 3.

\subsection{Single-subject decoding performance}

The performance of single-subject decoders is displayed in \cref{fig:increase_data}.
For the three participants with the largest datasets, comprising 13.5 hours of training data each (after validation/testing splits), we achieve an average top-10 accuracy of 27\%, with subject 3 reaching a maximum of 36\%.
For the remaining participants who had approximately 4 hours of training data, the average top-10 accuracy is 6\%, peaking at 9\% for subject 5.

To our knowledge, these results represent the first successful decoding of natural speech from fMRI data using a contrastive objective.
In addition, they demonstrate a clear positive correlation between the amount of training data available per individual and the decoding performance.
The decoding accuracy has not yet reached a plateau, indicating potential for further improvement with even larger acquisitions per participant.

Next, we investigate the impact of various components of our decoding pipeline on overall performance, as summarized in \cref{fig:setup}.
We start with a ``Base'' model, which consists of a simple Multi-Layer Perceptron (MLP) trained with a Mean Squared Error (MSE) loss function to predict BERT \citep{devlinBERTPretrainingDeep2019} text representations from fMRI data.
Subsequently, we iteratively incorporate modifications that enhance the decoder's performance.\\
First, to account for the haemodynamic response delay, we add a \textit{lag} of 6 seconds, thus predicting text representations from fMRI data acquired 6 seconds after the auditory stimulus.
Second, we enrich our text embeddings with context using a total \textit{context length} of 8 seconds,
Third, we smooth the fMRI signal by systematically averaging 4 consecutive brain volumes, potentially reducing noise and improving signal stability.
Then, we replaced BERT representations with text embeddings derived from LLM2Vec \citep{behnamghaderLLM2VecLargeLanguage2024}, a model specifically fine-tuned for sentence representation\footnote{We explored multiple models from the \href{https://huggingface.co/spaces/mteb/leaderboard}{Massive Text Emebdding Benchmark}. This is a very competitive benchmark comprising multiple sentence-level tasks which impose native token aggregation on the models. LLM2Vec is the model that produces the best representations for our decoding task.}, hypothesizing that these representations might be more suitable for our task.
We also transitioned from an MSE loss to a \textit{contrastive loss} objective, a common strategy in representation learning, to mitigate overfitting and improve generalization.
Finally, we replaced the simple MLP with a more complex Deep Neural Network architecture, termed ``Brain Decoder'', incorporating layer normalization and skip connections, inspired by \cite{scotti_mindeye2_2024}.
Each of these modifications incrementally improves the decoding accuracy, showing the cumulative benefit of these optimizations on the ability to decode natural speech from fMRI data.

\subsection{Impact of multi-subject decoding}

\begin{figure*}[!t]
    \centering
    \includegraphics[width=\textwidth]{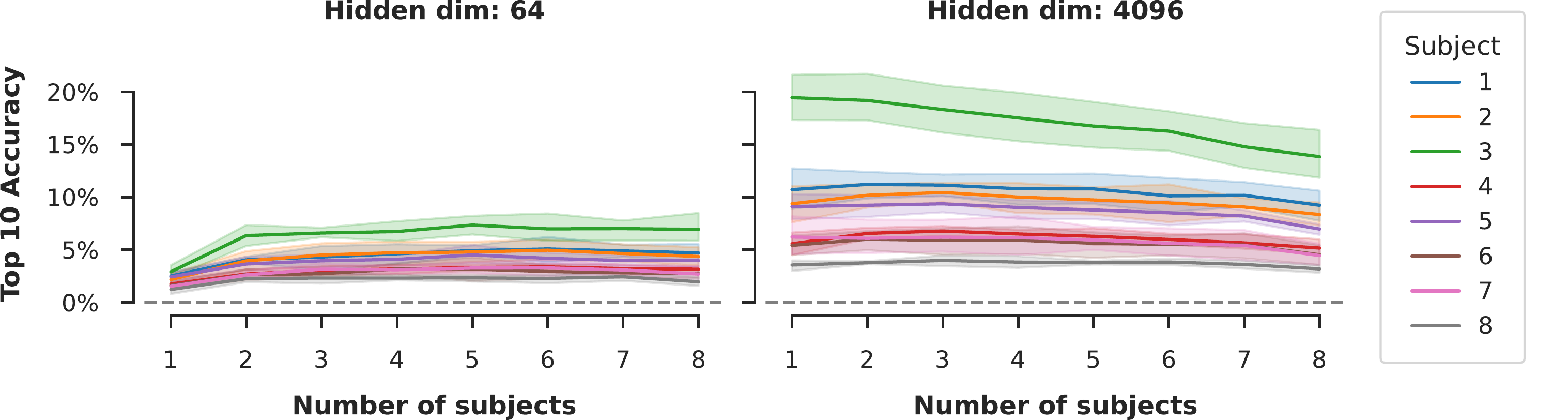}
    \vspace{0cm}
    \captionsetup{margin=5pt}
    \caption{\textbf{Impact of the number of subject used in the training set} Multi-subject decoders were trained with subject-specific layers for each of the 255 possible combinations of the 8 subjects. Then for each subject (color) and each number of subjects (x-axis), we display the best accuracy (y-axis) obtained with any of the combinations including this subject. We test \textit{small} decoders (left pane, hidden dimension 64) and large ones (right pane, hidden dimension 4096). Here we do not use the extra data available for the 3 first subjects.}
    \label{fig:increase_subjects}
\end{figure*}

\begin{figure*}[!t]
    \centering
    \vspace{0.1cm}
    \includegraphics[width=\textwidth]{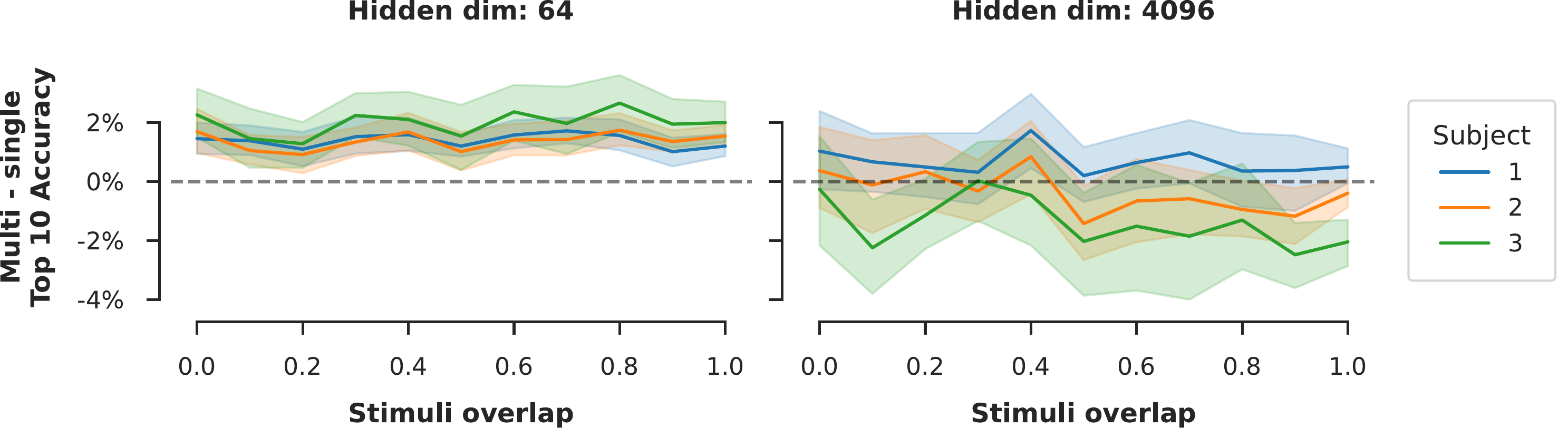}
    \vspace{0cm}
    \captionsetup{margin=5pt}
    \caption{
        \textbf{Impact of training stimuli  overlap} We train multi-subject decoders with subject-specific layers on subjects 1, 2 and 3 while varying the ratio of overlapping stimuli between the subjects.
        The graphics display the increment in accuracy over single-subject decoders ($y$ axis $=$ Multi $-$ Single top-10 accuracy), for \textit{small} decoders (left pane, hidden dimension 64) and large ones (right pane, hidden dimension 4096).
        For each of the 3 subjects we train on 1/3 of the available data ($\sim$~4.5~hours) for each overlap to be possible (in particular 0). The decoders are tested on the same stimuli, no matter the overlap.
    }
    \label{fig:overlap}
\end{figure*}

As illustrated in \cref{fig:increase_subjects}, multi-subject training does not improve individual-subject decoding performance.
It can even be detrimental for some participants, as the accuracy obtained with a multi-subject decoder is lower than the accuracy obtained with a single-subject decoder when the model size is large enough.
This effect is less pronounced for smaller models, but the improvement is still marginal.
We draw the following conclusion from this result: in the current data regime, there is evidence that models can learn mappings from brain signal to text embeddings that generalize to new stimuli acquired during new sessions, but that they do not yet model inter-subject variability.

Note that, in the previous multi-subject setup, the training, validation and test stories are the same across subjects.
Therefore, the diversity of the output space assessed by the decoder during training is the same as in the single-subject setup. Only the input space diversity is increased.
Our third experiment aims to investigate the impact of jointly increasing the input and output space diversity.\\
However, as illustrated in \cref{fig:overlap}, the overlap between the stimuli heard by different participants has a negligible effect on decoding performance in multi-subject setups.

\begin{figure}[!t]
    \centering
    \includegraphics[width=\textwidth]{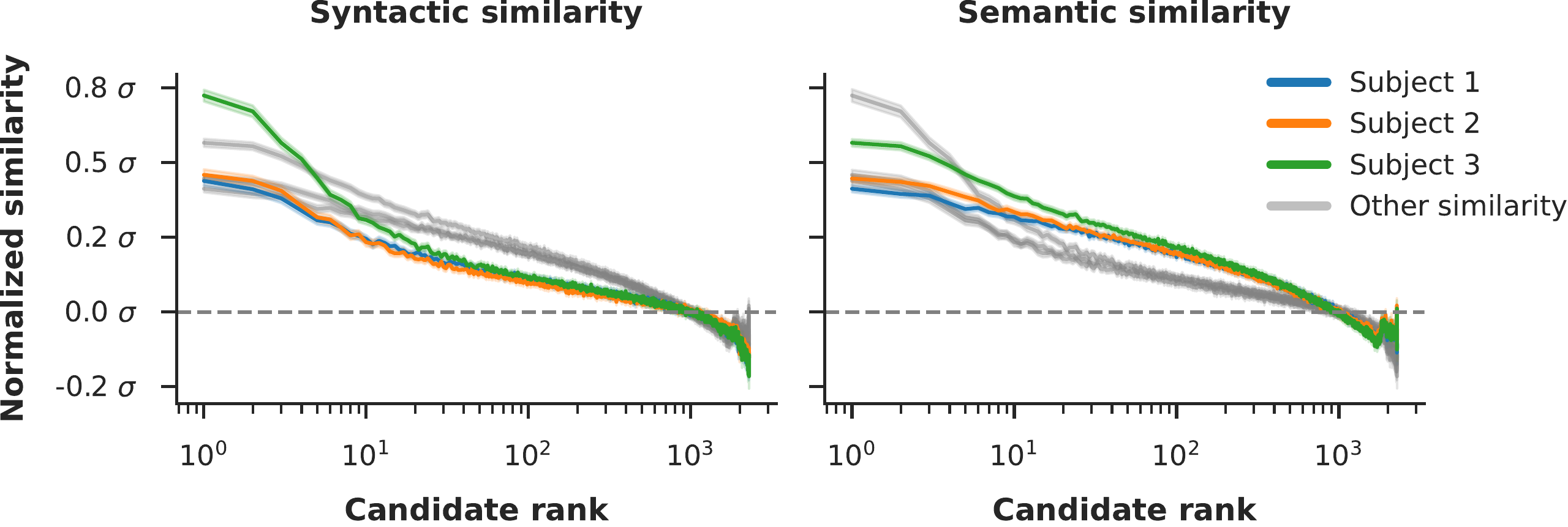}
    \vspace{0.2cm}
    \captionsetup{margin=5pt}
    \caption{
        \textbf{Profiles of average syntactic/semantic similarities}
        Syntactic (left) and semantic (right) similarities between the ground-truth text chunks and the $\sim$2k candidate chunks from a retrieval set sorted by decreasing cosine similarity of their representation to the predicted one from the decoder on the brain image.
        We normalize the similarities by subtracting the average and dividing by the standard deviation obtained by a baseline decoder producing random representations.
        Those results are from single-subject decoders trained on subjects 1, 2, and 3. Shaded areas indicate the 95\% confidence intervals computed over all time steps and stories from test splits.
    }
    \label{fig:nlp_similarities}
\end{figure}

\subsection{Impact of syntax/semantics}

Some stories are better decoded than others.
We investigate how the syntax and semantics of the stories impact decoding performance
through qualitative and quantitative analyses of the best and worst decoded stories.

\subsubsection{Qualitative analysis}
\begin{table*}[!ht]
    \centering
    \begin{tabular}{ p{0.125\linewidth} p{0.4\linewidth} p{0.4\linewidth} }
        \textbf{Group} & \textbf{10 best decoded stories} & \textbf{10 worst decoded stories} \\
        \toprule
        \textbf{Feature} & Personal experience & Ideas \& Reflection \\
        \midrule
        \textbf{Performance} & Top-10 accuracy:
        mean 51\%, min 44\%, max 59\% & Top-10 accuracy:
        mean 15\%, min 5\%, max 21\% \\
        \midrule
        \textbf{Narrative Voice} & Conversational, informal, direct address ("you know," "I mean") & More formal, polished, less direct address\\
        \midrule
        \textbf{Sentence Length} & Primarily short, choppy sentences for immediacy and emotional impact & Longer, more complex sentences reflecting intellectual exploration \\
        \midrule
        \textbf{Vocabulary} & Colloquial language, contractions, slang & Wider range of vocabulary, less colloquialism, more sophisticated phrasing \\
        \midrule
        \textbf{Use of Dialogue} & Frequent, natural-sounding dialogue to advance the narrative and express emotion & Dialogue used, but less prevalent and often serves an illustrative function \\
        \midrule
        \textbf{Sentence Structure} & Loose sentence structure with emphasis on emotion and immediacy & More complex sentence structures, often with clauses and sub-clauses \\
        \midrule
        \textbf{Tone} & Immediate, intimate, emotionally charged, often reflecting raw feelings and vulnerability & Reflective, analytical, measured, with a more intellectual distance and a focus on conveying insights \\
        \midrule
        \textbf{Descriptive Language} & Emphasis on sensory details and vivid imagery. & More use of metaphor and simile to add a higher level of detail and comparison to the text \\
        \midrule
        \textbf{Rhetorical Questions} & Frequent use of rhetorical questions & Less use of rhetorical questions \\
        \midrule
        \textbf{Emphasis} & Primarily to drive the narrative with a strong focus on the characters' internal experiences & Primarily to reflect, analyze and offer insights \\
        \midrule
        \textbf{Stories} & kiksuya, itsabox, thefreedomridersandme, comingofageondeathrow, hangtime, lifereimagined, cautioneating, thatthingonmyarm, fromboyhoodtofatherhood, threemonths & notontheusualtour, breakingupintheageofgoogle, jugglingandjesus, theadvancedbeginner, alternateithicatom, forgettingfear, avatar, theshower, treasureisland, bluehope\\
    \end{tabular}
    \vspace{0.4cm}
    \captionsetup{margin=5pt}
    \caption{
        \textbf{Qualitative analysis of the impact of semantics and syntax}
        We use Gemini \citep{reidGemini15Unlocking2024} to analyze the commonalities and differences between the 10 best and 10 worst performing stories in terms of semantics and syntax.
        Stories are sorted by the accuracy obtained from subject 3 and we display the mean/min/max accuracy for each group of stories.
    }
    \label{table:qualitative_analysis}
\end{table*}

Having identified the 10 worst and 10 best decoded stories from the performance on subject 3,
we fed their transcripts to Gemini \citep{reidGemini15Unlocking2024} to analyze their commonalities and differences.
Results are reported in \cref{table:qualitative_analysis}.
Overall, the best decoded stories are characterized by a more conversational and informal narrative voice, with direct address, colloquial language and simple sentences.
In contrast, the worst decoded stories are more formal, less direct in their address, with complex sentence structures and a wider range of vocabulary.

\subsubsection{Quantitative analysis}

We analyze syntactic and semantic similarities between the ground-truth text chunks and the $\sim$2k candidate chunks from a retrieval set.
Like previously in our retrieval setup, candidates are sorted by decreasing cosine similarity of their representation to the predicted one from a brain image with single-subject decoders trained on subjects 1, 2, or 3.
For each candidate, we compute a \textbf{syntactic similarity} ((1 - normalized Levenshtein distance) between POS-tagged chunks) and a \textbf{semantic similarity} (cosine similarity between GloVe embeddings of nouns, adjectives, and verbs as bag of words) with respect to the ground-truth chunk.
Similarities are normalized by subtracting the average and dividing by the standard deviation obtained from a decoder producing random representations, to remove intrinsic data contributions and ensure comparability.
We display the profile of both similarities, averaged over all stories and time steps, per candidate ranks for subjects 1, 2, and 3 in \cref{fig:nlp_similarities}.
The syntactic similarity profile starts slightly higher and decreases more sharply than the semantic one, indicating that the decoder better differentiates between syntactically dissimilar chunks than semantically dissimilar ones.

\section{Discussion}

Our findings provide several key insights into the optimization of fMRI data acquisition for decoding perceived natural speech, particularly when working with a limited number of participants.
First, we have demonstrated the feasibility of accurately predicting LLM-derived text representations from fMRI activity using a contrastive learning objective. This extends previous work on natural language decoding \citep{willettHighperformanceSpeechNeuroprosthesis2023,tang_semantic_2023,defossez_decoding_2022} by systematically exploring the impact of key methodological choices, such as context length, lag, and fMRI smoothing.
The incremental improvements observed in Figure \ref{fig:setup} highlight the impact of these parameters on decoding performance.
\newpage
The most striking result is perhaps the lack of improvement when transitioning from the single-subject to the multi-subject training approach (Figure \ref{fig:increase_subjects}).
This contrasts with some findings in other neuroimaging domains \citep{defossez_decoding_2022,dascoliDecodingIndividualWords2024,mentzelopoulosNeuralDecodingStereotactic2024} and even within the domain of fMRI \citep{thual_aligning_2023,aggarwalAcrosssubjectEnsemblelearningAlleviates2024}, where multi-subject training often leads to slightly better generalization.
Interestingly, \cite{tangSemanticLanguageDecoding2025} shows slight improvements in multi-subject setups compared to single-subject ones for a language decoding task on the LeBel dataset. However, their approach produces text using a beam search guided by a brain encoder. We hypothesize that gains reported in the multi-subject setup may come from the fact that functional alignment smooths data in a way that improves the encoder's stability. Our decoding-first approach is immune from this specific drawback.\\
Our results suggest that, at least within the current data regime -- a relatively small cohort of 8 individuals with extensive data for only 3 of them -- inter-subject variability in the neural encoding of natural speech is substantial.
The subject-specific layer (SSL) approach, while theoretically promising, did not fully overcome this challenge.
Also, simply increasing the number of subjects, without accounting for individual differences in brain organization and function, may not be sufficient to boost decoding accuracy. It may even be detrimental, especially for larger models that can potentially overfit to subject-specific noise in the combined dataset.

The negligible impact of stimulus overlap (Figure~\ref{fig:overlap}) further supports the notion that inter-subject variability, rather than stimulus diversity, is the primary limiting factor in multi-subject decoding.
While it is intuitively appealing to train a decoder on shared stimuli to facilitate alignment across subjects, our results indicate that this has little beneficial effect. This finding has practical implications for experimental design: researchers can prioritize collecting diverse and ecologically valid stimuli for each participant, rather than being constrained by the need for extensive stimulus overlap.
This also suggests that different datasets can be combined, even if participants did not listen to the same stimuli, or even stimuli in different languages, provided that multi-lingual LLMs are used to derive language-agnostic text representations.

The qualitative and quantitative analyses of best- versus worst-decoded stories (Table \ref{table:qualitative_analysis} and Figure \ref{fig:nlp_similarities}) reveal intriguing differences in the linguistic features that drive decoding performance.
The decoder appears to be more sensitive to syntactic structure than to semantic content, as evidenced by the steeper decline in syntactic similarity compared to semantic similarity.
Furthermore, stories with more complex syntax and richer semantic content (e.g., abstract ideas, metaphorical language) proved more challenging to decode.
This suggests that current decoding approaches, while effective, may still be biased towards capturing more superficial aspects of language processing.
This finding is consistent with results reported in \cite{tuckuteLanguageBrainsMinds2024}, in which the authors show that different types of sentences can either drive or suppress activations in the language network, and that syntax has a high impact in that regard.
Future work should explore methods that are more robust to variations in linguistic complexity and that can better capture the nuances of meaning conveyed in natural speech.

Taken together, our results suggest that a "deep phenotyping" strategy, where extensive data is collected per participant, is more effective than collecting less data from a larger number of subjects, at least in the current data regime.
This does not preclude the possibility that multi-subject training could become beneficial with a much larger cohort -- e.g., hundreds of participants -- or with more sophisticated methods for modeling inter-subject variability, such as advanced functional alignment techniques \citep{haxby_common_2011,thualAligningIndividualBrains2022} or hierarchical Bayesian models \citep{gelmanBayesianDataAnalysis2013}.
However, collecting deeply phenotyped data from such a large number of participants would require considerable resources.

\section{Limitations}

First, unlike \cite{tang_semantic_2023} and \cite{ye2025generative}, we do not reconstruct actual text from brain activity.
This could in part be alleviated by predicting text embeddings which can be used to condition the sampling process of a pre-trained language model \citep[see e.g.][]{mokady_clipcap_2021,ye2025generative}.

More generally, we acknowledge that the current decoding problem is still empirically ill-posed: contrary to \cite{dascoliDecodingIndividualWords2024} who used MEG and EEG data, the low temporal resolution of fMRI makes it challenging to decode information at the word level. However, as illustrated with varying context lengths, the nature and quality of embeddings used to train the decoder greatly impact its performance.

Both of these reasons make it impossible to find a straightforward way to compare our results to that of previously published language-decoding papers.
Note that we refrain from using the predicted embeddings as inputs for text-generative models as the resulting downstream evaluation metrics would highly depend on the generative model at hand, and less explicitly on the decoding model itself.

Lastly, the main limiting factor of our work lies in the amount of data used for training.
Our findings are valid for a small cohort of highly phenotyped participants, but may not be for larger cohorts, especially regarding the impact of stimuli overlap across participants.
We advocate that acquiring larger datasets is the way forward.

\onecolumn

\bibliography{main}

\appendix

\section{Supplementary Materials}

\begin{table}[ht]
    \centering
    \begin{tabular}{ p{0.225\linewidth} p{0.1\linewidth} p{0.55\linewidth} }
        \toprule
        \textbf{Hyperparameter} & \textbf{Value} & \textbf{Description} \\
        \\
        \midrule
        \multicolumn{3}{c}{\textbf{Data Configuration}} \\
        \midrule
        Context Length & 3 & Number of preceding text chunks concatenated (= 6s) \\
        Temporal Smoothing & 4 & Number of preceding brain volumes averaged (= 8s) \\
        Lag (s) & 6 & Delay between a brain volume and the target chunk \\
        Top Encoding Voxels & 4096 & Number of voxels selected based on encoding performance \\
        \\
        \midrule
        \multicolumn{3}{c}{\textbf{Brain Decoder Configuration}} \\
        \midrule
        Hidden Dimension & 64 or 4096 & Size of the hidden layers (both are considered in the experiments) \\
        Number of Linear Layers & 1 & \\
        Number of Residual Layers & 1 & \\
        Total number of layers & 3 & Subject specific layer + Linear + Residual\\
        Normalization Type & Layer & \\
        Activation Function & GELU & \\
        \\
        \midrule
        \multicolumn{3}{c}{\textbf{Training Configuration}} \\
        \midrule
        Temperature & 0.7 & Temperature of the contrastive loss \\
        Learning Rate & 1e-4 & \\
        Weight Decay & 5e-4 & \\
        Patience & 20 & Number of epochs with no improvement before early stopping \\
        Scheduler Patience & 5 & Number of epochs with no improvement before reducing the learning rate \\
        Scheduler Factor & 0.5 & Factor by which the learning rate is reduced \\
        Batch Size & 1 & Number of stories per batch \\
        Max Epochs & 200 & \\
        \bottomrule
        \\
    \end{tabular}
    \vspace{0.4cm}
    \captionsetup{margin=5pt}
    \caption{
        \textbf{Hyperparameters of the DNNs} Comprehensive list of the hyperparameters used in the study. They were determined through a Bayesian grid search maximizing cross-validated top-10 accuracy on subjects 1, 2, and 3 in the multi-subject setup. In this setup small decoders (hidden dimension 64) have 250k parameters in their backbone and 250k (per subject) in the subject-specific layers. Large decoders (hidden dimension 4096) have 70e6 in addition to 17e6 (per subject) parameters.
    }
    \label{table:hyperparameters}
\end{table}

\end{document}